\documentclass[aps,twocolumn,showpacs,superscriptaddress,prb,reprint,floatfix]{revtex4-1}

\usepackage{amsmath}
\usepackage{amsfonts}
\usepackage{amssymb}
\usepackage{graphicx}
\usepackage{bm}
\usepackage[binary,amssymb]{SIunits}

\usepackage{color}
\definecolor{Blue}{rgb}{0.3,0.3,0.9}
\definecolor{Red}{rgb}{0.94,0.14,0}
\definecolor{Green}{rgb}{0.3,0.6,0.3}

\begin{document}

\title{Spin-polarized electric current in silicene nanoribbons induced by atomic adsorption}

\author{C.\ N\'{u}\~{n}ez}

\affiliation{Departamento de F\'{\i}sica, Universidad T\'{e}cnica 
Federico Santa Mar\'{\i}a, Casilla 110 V, Valpara\'{\i}so, Chile}

\affiliation{Department of Physics, University of Warwick, Coventry, 
CV4 7AL, United Kingdom}

\author{P.\ A.\ Orellana}

\affiliation{Departamento de F\'{\i}sica, Universidad T\'{e}cnica 
Federico Santa Mar\'{\i}a, Casilla 110 V, Valpara\'{\i}so, Chile}

\author{L.\ Rosales}

\affiliation{Departamento de F\'{\i}sica, Universidad T\'{e}cnica 
Federico Santa Mar\'{\i}a, Casilla 110 V, Valpara\'{\i}so, Chile}

\author{R.\ A.\ R\"{o}mer}

\affiliation{Department of Physics, University of Warwick, Coventry, 
CV4 7AL, United Kingdom}

\author{F.\ Dom\'{i}nguez-Adame}

\affiliation{GISC, Departamento de F\'{\i}sica de Materiales, Universidad
Complutense, E-28040 Madrid, Spain}

\affiliation{Department of Physics, University of Warwick, Coventry, 
CV4 7AL, United Kingdom}

\begin{abstract}

We investigate the non-equilibrium transport properties of a silicene armchair nanoribbon with a random distribution of adsorbed atoms in apex positions. A ferromagnetic insulator grown below the nanoribbon splits spin-up and spin-down electron bands and gives rise to a spin polarization of the conductance. The conductance vanishes when the Fermi energy matches the adatom levels due to the coupling of adatom localized states with the continuum spectra of the nanoribbon. This is the well-known Fano effect, resulting in a spin-dependent anti-resonance in the conductance. The different anti-resonance energies of spin-up and spin-down electrons give rise to a full spin polarization of the conductance in a broad energy window. This spin-dependent Fano effect opens the possibility to using it in spintronics as a tuneable source of polarized electrons.

\end{abstract}


\maketitle

\section{Introduction}

 The field of spintronics is rapidly developing from its roots in magnetic metal multilayers. In recent years, two-dimensional (2D) materials came to the forefront in spintronics and advances in this field are expected to occur based on hybrid systems.\cite{Wei16a} Although graphene at present dominates the sector due to its high-electron mobility, a plethora of other novel 2D materials offers fascinating fundamental properties for spin transport and controlled spin-light interaction.\cite{Fert08,Zabel09,Joshi16} In this context, silicene is a particularly promising candidate for the design of spintronic devices.\cite{Wang15a,Zhao16} The two sublattices of the honeycomb lattice in silicene are not coplanar and first-principles calculations suggest that the spin-orbit interaction opens a sizable gap at the Dirac point of the order of~$\unit{1.55}\milli\electronvolt$.\cite{Liu11} 

Half-metals, in which one spin channel is conductive but the other one is insulating or semiconducting, turn out to be a key ingredient to achieve spin polarized currents. Hybrid structures of 2D materials and ferromagnetic insulators, like EuO, EuS, yttrium iron garnet or cobalt ferrite, provide a route to induce half-metallicity\cite{Haugen08,Lee10,Yang13,Wang15b,Mendes15,Wei16b,Leutenantsmeyer17} and pave the way for spintronic applications.\cite{Semenov07,Michetti11,Munarriz12,Munarriz13,Qiao14,Diaz14,Pietrobon15,Saiz-Bretin15,Lee16,Myoung16} The ferromagnetic insulator induces a proximity exchange interaction between the spins in the magnetic and non-magnetic material that results in a spin modulation without compromising the crystallinity of the structures.\cite{Swartz12}

Recently we have proposed a novel spin-filter device based on a silicene nanoribbon.\cite{Nunez16} A ferromagnetic insulator below the nanoribbon gives rise to the spin polarization of the electric current. A random distribution of vacancies causes Anderson localization of electrons. Since the localization length was found to be spin-dependent, only electrons with one spin orientation can reach the drain contact because their localization length is larger than the length of the device while electrons with opposite spin are largely back-reflected.\cite{Nunez16} Besides vacancies, other types of point defects can affect electron transport in silicene nanoribbons. In particular, metal adatoms present much stronger binding to silicene than to graphene\cite{Lin12} and can induce a transition from semimetallic to semiconducting behavior\cite{Ersana14,Du14} or produce a quantum anomalous Hall effect.\cite{Zhang13} Moreover, chemisorption of a single H atom on silicene induces the formation of a localized state around the adatom which acts as a resonant scatterer for charge carriers.\cite{Pizzochero16} Hence, a significant reduction of the electron mobility is anticipated  since the absence of clustering prevents the conversion of isolated adatoms into clusters, which are known to have a much smaller effect on electron mobility.\cite{Pizzochero16} 

Theoretical methods and modeling are needed to understand the role of adatoms in the electron transport of silicene and their influence on the spin filtering capabilities induced by the proximity exchange interaction. In this paper we address the effects of a random distribution of adatoms on the electron transport properties of narrow silicene nanoribbons. Interestingly, De Padova \emph{et al.}\cite{Padova10} have already synthesized silicene nanoribbons with very large aspect ratio (several nanometres in length and a constant width about $\unit{2}\nano\metre$). These nanoribbons usually display higher electrical sensitivity to the adsorption of certain molecules, such as CO, at low concentration compared to graphene nanoribbons.\cite{Osborn14}  The tunnel coupling between adatoms and silicon atoms induces an electronic Fano effect\cite{Fano61} that makes the conductance vanish when the Fermi level matches the resonant energy induced by the adatoms. The resonant energy turns out to be independent of the random distribution of adatoms, provided that they do not cluster. When the nanoribbon is in close proximity  to a ferromagnetic insulator, the resonant energy depends on the electron spin and consequently the electric current can be highly spin-polarized. Our results expand the base of available materials to designing a tuneable source of polarized electrons for spintronics.

\section{Theoretical model}

\begin{figure}[tb]
\centering
\includegraphics[clip,width=0.80\columnwidth,angle=0,clip]{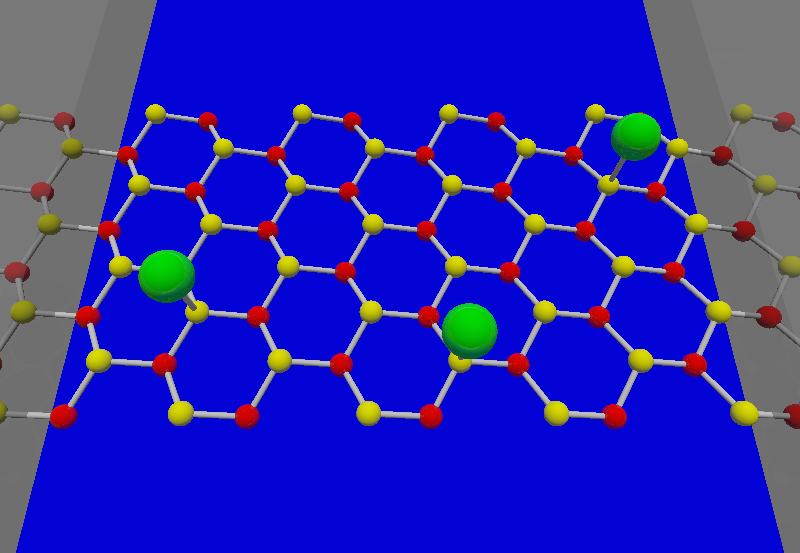}
\caption{Schematic view of the device. The armchair silicene nanoribbon is connected to left~(L) and right~(R) leads, with a ferromagnetic insulator grown below it (shown as the blue region in the figure). A random distribution of adatoms is shown as green spheres. Smaller yellow and red spheres indicate the non-equivalent Si atoms in the silicene nanoribbon.}
\label{fig1}
\end{figure}

The system consists of a narrow silicene nanoribbon of width $W$ and length $L$ connected to source and drain leads, as shown schematically in Fig.~\ref{fig1}. In order to avoid topologically protected edge states that appear at the Fermi energy in zigzag nanoribbons,\cite{Brey06} we restrict ourselves to nanoribbons with armchair edges hereafter. A ferromagnetic insulator below the nanoribbon induces a spin-splitting of the electronic states. Consequently, electrons in the nanoribbon will be subject to a positive or negative constant potential, according to their spin.

Electrons in the silicene honeycomb lattice are described by a single $\pi$-orbital tight-binding Hamiltonian of the form $H = H_\mathrm{SN}+H_\mathrm{ad}+H_\mathrm{tun}+H_\mathrm{exc}$. The Hamiltonian for the electron in the silicene nanoribbon is\cite{Ezawa12}
\begin{equation}
H_\mathrm{SN}=-t \sum_{\langle i,j \rangle\sigma}c_{i\sigma}^{\dag}c_{j\sigma}^{} 
+ \frac{\lambda_\mathrm{SO}}{3 \sqrt{3}}
\sum_{\langle\langle i,j\rangle\rangle\sigma \tau} 
\nu_{ij}  c_{i\sigma}^{\dag} \sigma_z c_{j\tau}^{} \ ,
\label{eq:01}
\end{equation}
where $c_{i\sigma}^{\dag}$ ($c_{i\sigma}^{}$) creates (annihilates) an electron at silicon atom $i$. Sums over $\langle i,j \rangle$ and $\langle\langle i,j\rangle\rangle$ run over nearest and next-nearest neighbour sites, respectively. The spin indices $\uparrow, \downarrow$ are indicated by $\sigma$ and $\tau$ hereafter. The first term corresponds to the nearest neighbour hopping energy $t=\unit{1.6}\electronvolt$ and the second term represents the spin-orbit coupling with $\lambda_\mathrm{SO}=\unit{3.9}\milli\electronvolt$, where $\nu_{ij}=\pm 1$ is the Haldane factor\cite{Haldane88} and $\sigma_z$ is the Pauli matrix.
The Hamiltonian corresponding to adatoms levels is given as
\begin{equation}
H_\mathrm{ad}=\sum_{{j\in \mathcal{L}},\sigma}\varepsilon_\mathrm{ad}^{} d_{j\sigma}^{\dag}d_{j\sigma}^{} \ ,
\label{eq:02}
\end{equation}
where $d_{j\sigma}^{\dag}$ ($d_{j\sigma}^{}$) creates (annihilates) an electron at the adatom $j$, with the index $j$ running over those silicon sites with an attached adatom, denoted by $\mathcal{L}$ above. Adatoms are assumed to occupy apex positions and electrons may tunnel from and to the silicon atom on which the adatom is located.  Therefore, the tunnel coupling is expressed as
\begin{equation}
H_\mathrm{tun}=t_\mathrm{ad}\sum_{j\in \mathcal{L},\sigma}(c_{j\sigma}^{\dag}d_{j\sigma}^{}
+d_{j\sigma}^{\dag}c_{j\sigma}^{})\ .
\label{eq:03}
\end{equation}
where $t_\mathrm{ad}$ is the hopping parameter between the adatom and the silicene nanoribbon.
Finally, the term
\begin{equation}
H_\mathrm{exc}=\sum_{i\sigma}M c_{i\sigma}^{\dag}\sigma_z c_{i\sigma}^{}
\label{eq:04}
\end{equation}
describes the spin-splitting of electron states due to the proximity exchange interaction with the ferromagnet. 
It raises (lowers) the energy levels of spin up (spin down) electrons by an amount $+M$ ($-M$).\cite{Nunez16}

We study electron transport across the nanoribbon using the Green's function formalism combined with decimation techniques.\cite{Lambert80,Datta95} This approach allows us to obtain the transmission coefficient $T_\sigma(E)$ for an electron with energy $E$ and spin $\sigma$. Details of the calculations can be found in Ref.~\onlinecite{Nunez16}. In the linear response regime, the conductance is calculated from the transmission coefficient using the Landauer formula at zero temperature\cite{Landauer70} 
\begin{equation}
I_\sigma(V,E_F) = \frac{e}{\pi \hbar}\int_{E_F - eV/2}^{E_F +eV/2} T_\sigma(E) dE \ ,
\label{eq:05}
\end{equation}
where $E_F$ is the equilibrium Fermi energy. For the sake of simplicity, here we assume that the voltage drops across the conductor-electrode interfaces only, although this assumption does not affect significantly the current-voltage characteristics.\cite{Ojeda12} The total polarization of the spin-dependent linear conductance $G_\sigma(E_F)=I_\sigma(V,E_F)/V$ ($V\to 0$) is defined as
\begin{equation}
P(E_F) = \frac{G_\uparrow(E_F)-G_\downarrow(E_F)}{G_\uparrow(E_F)
+G_\downarrow(E_F)}\ ,
\label{eq:06}
\end{equation}
and it will be the figure of merit to assess the spin filtering properties of the device. 

\section{Results}

\subsection{Linear conductance in the absence of ferromagnet}

Silicene nanoribbons, grown in a controlled environment of gas such as hydrogen, oxygen, boron, lithium or silver, will be covered by a random distribution of adatoms while possibly retaining their honeycomb structure.\cite{Du14} Model parameters as the energy level $\varepsilon_\mathrm{ad}$ and the tunnel energy $t_\mathrm{ad}$ will depend on the particular species adsorbed by the nanoribbon. In addition, the fraction $c$ of silicon atoms with an attached adatom will vary according to the growth conditions. Another crucial parameter of the model is the spin splitting due to proximity exchange interaction with the ferromagnetic insulator. In our simulations we take typical values of these magnitudes to illustrate the feasibility of the proposed device. Other values of the model parameters do not qualitatively change our main conclusions. 

\begin{figure}[tb]
\centerline{\includegraphics[width=0.95\columnwidth]{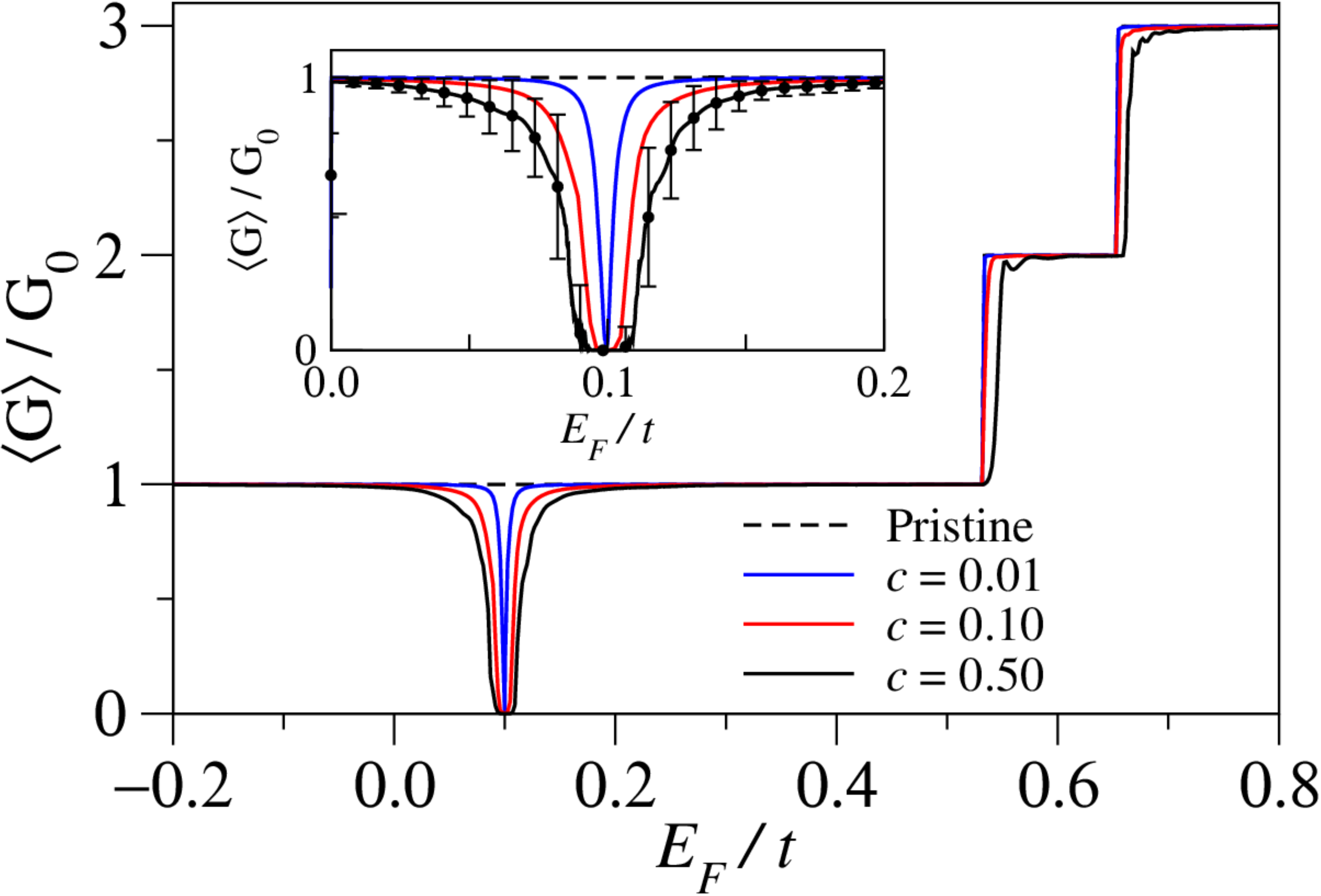}}
\caption{(Color online) Average conductance as a function of the Fermi energy of a silicene nanoribbon when $M=0$. Solid lines correspond to the average over $100$ realizations of random samples with adatom concentration $c=0$ (pristine), $0.01$, $0.10$ and $0.50$. The energy level of the adatom is $\varepsilon_\mathrm{ad}=0.10t$ and the hopping parameter $t_\mathrm{ad}=0.10t$ with $t=\unit{1.6}\electronvolt$. The inset shows an enlarged view of the anti-resonance. Only the error bars of the more disordered sample ($c=0.5$) are displayed for clarity. Fluctuations strongly diminish upon lowering $c$.}
\label{fig2}
\end{figure}

In Fig.~\ref{fig2} we show the average conductance $\langle G\rangle/G_0$ as a function of the Fermi energy $E_F$ when the energy level of the adatom is $\varepsilon_\mathrm{ad}=0.10t$ ($\unit{0.16}\electronvolt$) and the hopping parameter $t_\mathrm{ad}=0.10t$ ($\unit{0.16}\electronvolt$). Here $G_0=e^2/h$ is the quantum of conductance per spin. Spin-splitting effects are not considered for the moment ($M=0$). We set the size of the system of the nanoribbons, $W\times L$, as width $W=\unit{2.35}\nano\metre$ and length $L=\unit{23.22}\nano\metre$ that were used in all of our calculations. Results for the pristine sample ($c=0$) are compared to the average over $100$ realizations of random samples with concentrations from $c=0.01$ up to $c=0.50$. The characteristic quantum plateaus of the conductance are clearly revealed when $c=0$ but the conductance drops abruptly at the adatom energy at finite values of $c$. The occurrence of an anti-resonance in the conductance can be traced back to quantum interference between the states in the continuum of the nanoribbon and the localized states of the adatoms. This is nothing but the electronic analogue of the optical Fano effect.\cite{Fano61} 
It originates from the interference of two coexisting paths for a traveling electron in the system. One path is a direct way that traverses the nanoribbon while the second path includes a hopping on and off the adatom and then the electron continues with propagation. The destructive interference between these two paths is at the heart of the Fano anti-resonance. The conductance around a Fano anti-resonance at an energy $E_\mathrm{ar}$ can be fitted by the general expression $G(E_F)/G_0 \sim (\widetilde{\epsilon} + q)^2/(\widetilde{\epsilon\,}^{2}+1)$. Here $\widetilde{\epsilon} = (E_F-E_\mathrm{ar})/\gamma$ corresponds to the normalized energy of the Fano anti-resonance, $q$ is an adjustable parameter related to the phase shift originated in the interference phenomena and $\gamma$ is an effective coupling between the adatom and the nanoribbon.\cite{Rezapour14,Petrovic15}

When the concentration of adatoms increases the anti-resonance does not shift but the dip becomes broader. For high concentration of adsorbed atoms a gap centered at the adatom energy opens. The anti-resonance remains despite having high concentrations of adatoms and the fully covered nanoribbon ($c = 1$) displays a gap of width $t_\mathrm{ad}^2/t$ centered at the adatom energy level, as can be demonstrated as follows. The dispersion relation in the pristine nanoribbon is
\begin{equation}
E_{0}(k_{\parallel},k_{\bot})=\pm t \sqrt{1+4\cos k_{\parallel} \cos k_{\bot} + 4 \cos k_{\bot}^{2}}\ ,
\label{eq:07}
\end{equation} 
where $k_{\parallel}$ and $k_{\bot}$ are the longitudinal and transverse wave numbers
in units of the inverse of the lattice period,
respectively. 
Here the subscript '$0$' refers to the absence of adatoms.
The energy spectrum is symmetric about $E=0$ and the two bands touch at this energy if $k_{\bot}=2\pi/3$. Then the nanoribbon is metallic and the longitudinal dispersion relation is
\begin{equation}
\varepsilon_{0}(k_{\parallel})\equiv E_{0}(k_{\parallel},k_{\bot}=2\pi/3)=\pm 2t \left|\sin(k_{\parallel}/2)\right|\ .
\label{eq:08}
\end{equation}

On the other side, in the fully covered nanoribbon ($c=1$) the energy levels of the Si atoms are renormalized and become energy-dependent after the substitution $\varepsilon_\mathrm{Si} \to \varepsilon_\mathrm{Si} + t_\mathrm{ad}^2/\left[\varepsilon(k_{\parallel})-\varepsilon_\mathrm{ad}\right]$. Now $\varepsilon(k_{\parallel})$ stands for the dispersion relation when $c=1$. Notice that we have taken $\varepsilon_\mathrm{Si}=0$ throughout this work. Introducing this substitution in Eq.~(\ref{eq:08}) yields

\begin{equation}
\varepsilon(k_{\parallel})=\frac{t_\mathrm{ad}^2}{\varepsilon(k_{\parallel})-\varepsilon_\mathrm{ad}}
\pm 2t \left|\sin(k_{\parallel}/2)\right|\ .
\label{eq:09}
\end{equation}
In this case, a gap opens at $k_{\parallel}=\pi$. Taking $\varepsilon_{ad}=0$ for simplicity, the two roots of Eq.~(\ref{eq:09}) at $k_{\parallel}=\pi$ are found to be $\varepsilon_{\pm}=\pm t \mp t\sqrt{1+(t_\mathrm{ad}/t)^2)}$. Consequently, the magnitude of the gap is $\left|\varepsilon_{+}-\varepsilon_{-}\right|=2\sqrt{t^2+t^{2}_\mathrm{ad}}-2t \simeq t^{2}_\mathrm{ad}/t$, where we have taken into account that $t_\mathrm{ad}\ll t$.

\begin{figure}[tb]
\centering
\includegraphics[width=0.95\columnwidth]{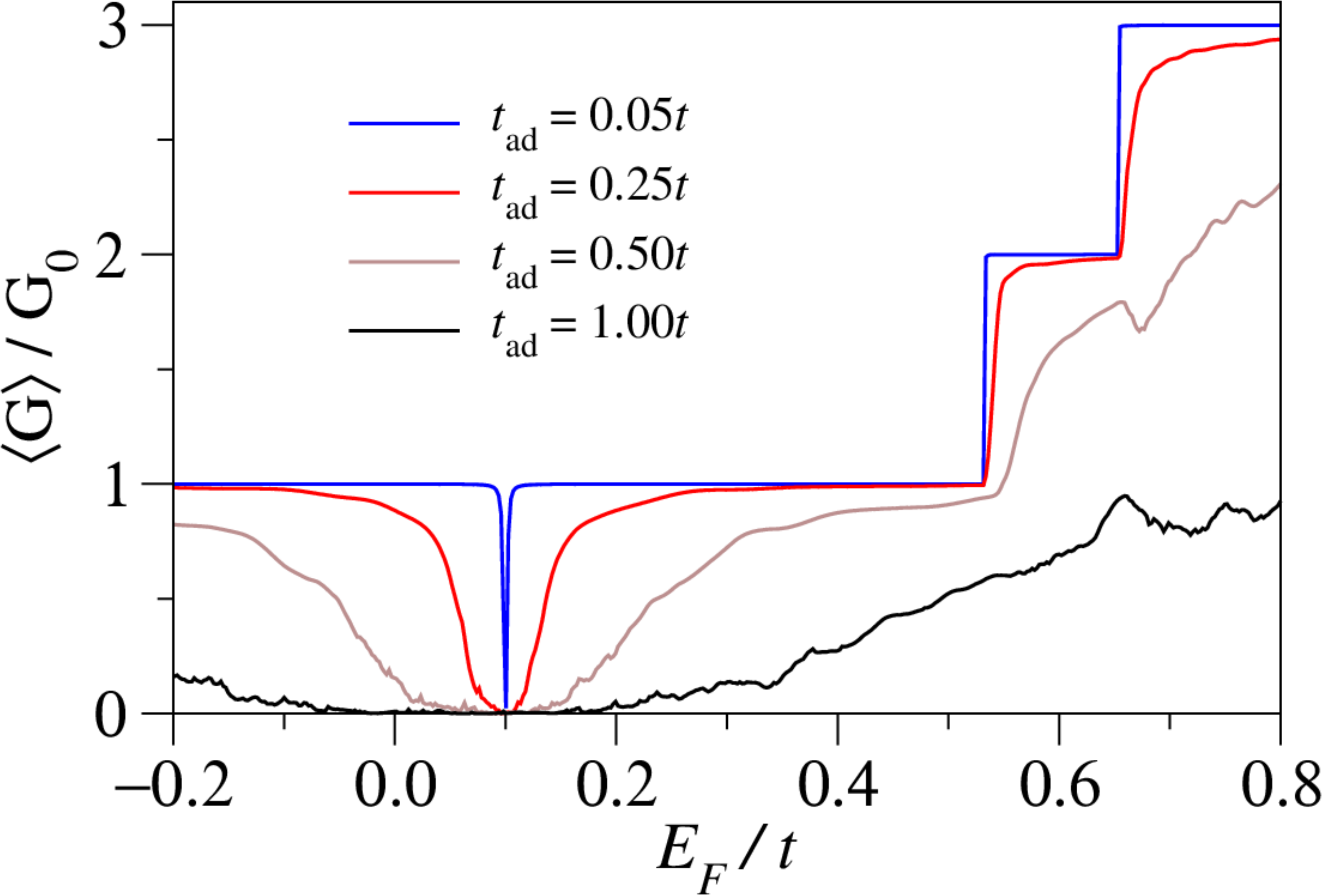}
\caption{(Color online) Average conductance as a function of the Fermi energy of a silicene nanoribbon when $M=0$. Results were averaged over $100$ realizations of random samples with adatom concentration $c=0.05$ and $t_\mathrm{ad}=0.05t$, $0.25t$, $0.50t$ and $1.00t$, where $t=\unit{1.6}\electronvolt$. The energy level of the adatom is $\varepsilon_\mathrm{ad}=0.10t$.}
\label{fig3}
\end{figure}

We now turn to the dependence of the anti-resonance on the model parameters $\varepsilon_\mathrm{ad}$ and $t_\mathrm{ad}$. Figure~\ref{fig3} displays the average conductance $\langle G\rangle$ as a function of the Fermi energy $E_F$ in units of the hopping energy $t$ for different values of $t_\mathrm{ad}$ when $\varepsilon_\mathrm{ad}=0.10t$. Results were averaged over $100$ realizations of random samples with adatom concentration $c=0.05$. The average conductance vanishes at $E_F=\varepsilon_\mathrm{ad}$ and the anti-resonance becomes broader on increasing $t_\mathrm{ad}$. It can be shown that the width of the anti-resonance scales quadratically with $t_\mathrm{ad}$, provided that the adatom concentration is not large. We have also performed calculations when the hopping energy 

varies at random with mean value $t_\mathrm{ad}$ from adatom to adatom

(not shown in the figure). The general trend is the same as before, in the sense that the larger the fluctuation of the hopping energy, the wider the anti-resonance. But even if this hopping is random, the conductance vanishes at $E_F=\varepsilon_\mathrm{ad}$. Therefore, we come to the conclusion that the Fano anti-resonance effect is a very robust phenomenon, which is advantageous for applications.

\subsection{Electric current in the absence of ferromagnet}

\begin{figure}[tb]
\centering
\includegraphics[width=0.95\columnwidth]{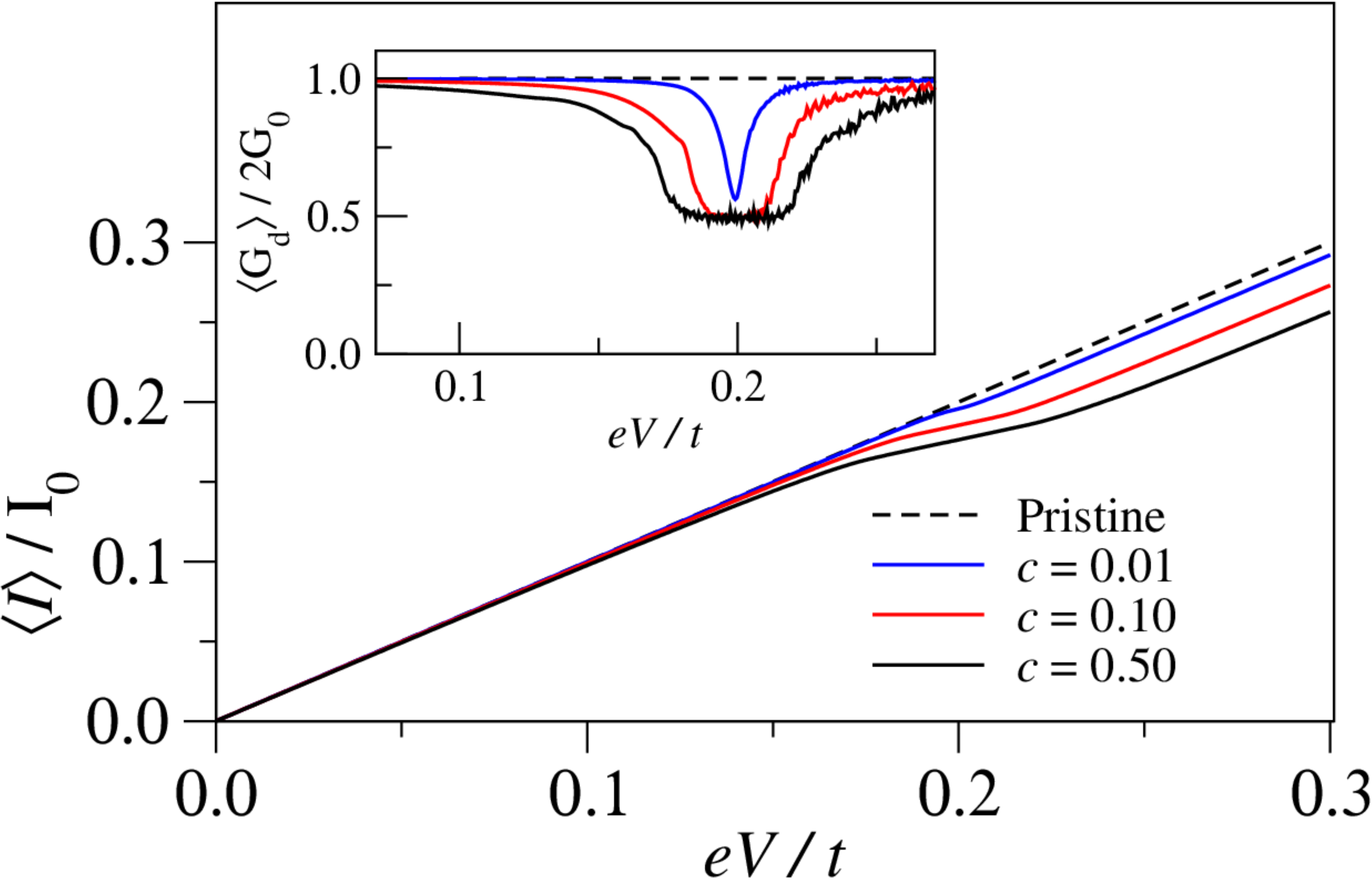}
\caption{(Color online) Electric current through silicene nanoribbons, averaged over $100$ realizations of random samples, in units of $I_0=et/\pi\hbar$, as a function of the source-drain potential energy drop $eV$, for $E_F=0$ and the rest of parameters as in Fig.~\protect{\ref{fig2}}. The inset shows the corresponding average differential conductance $\langle G_\mathrm{d}\rangle = \langle dI/dV\rangle$, expressed in units of the quantum of conductance $2G_0=2e^2/h$.}
\label{fig4}
\end{figure}

Figure~\ref{fig4} displays the current-voltage characteristics for different values of the adatom concentration when the Fermi energy at equilibrium is $E_F=0$. The rest of the parameters are the same as in Fig.~\ref{fig2}. The pristine nanoribbon displays a perfectly ohmic current-voltage characteristics over the entire range of voltage~$V$. However, the response becomes non-ohmic at finite adatom concentration and an inflection point appears at $eV=2\varepsilon_\mathrm{ad}=2t$. The point reveals itself as a minimum in the differential conductance, as seen in the inset of Fig.~\ref{fig4}. 

The drop of about $50\%$ of the quantum of conductance $2G_0$ can be easily understood as follows. Assuming that the transmission coefficient does not change much with the applied voltage, one can obtain a simple expression for the differential conductance from the electric current~(\ref{eq:05}). The result is $\langle G_d \rangle/ 2G_0 \simeq (1/2) \left[ T_{\sigma}(E_F + eV/2) + T_{\sigma}(E_F - eV/2)\right]$. When $E_F = 0$ (as in the inset of Fig.~\ref{fig4}), the last term equals unity since the Fano anti-resonance lies on the positive-energy side and the nanoribbon is metallic. Thus $\langle G_d \rangle/ 2G_0 \simeq (1/2) \left[ 1+T_{\sigma}(eV/2)\right]$. When $eV/2$ is close to the Fano anti-resonance, located at an energy $\varepsilon_\mathrm{ad} = 0.1$ (i.e. $eV \sim 0.2$), the transmission vanishes and $\langle G_d \rangle/ 2G_0 \simeq 1/2$. On the contrary, far from the Fano anti-resonance, the transmission becomes unity and $\langle G_d \rangle/ 2G_0 = 1$. 

Also notice that the gap increases with the adatom concentration, as we anticipated above. Hence, atomic adsorption induces a chemically-tunable gap in silicene nanoribbons.

\subsection{Polarization effects of the ferromagnet}

\begin{figure}[tb]
\centering
\includegraphics[width=0.95\columnwidth]{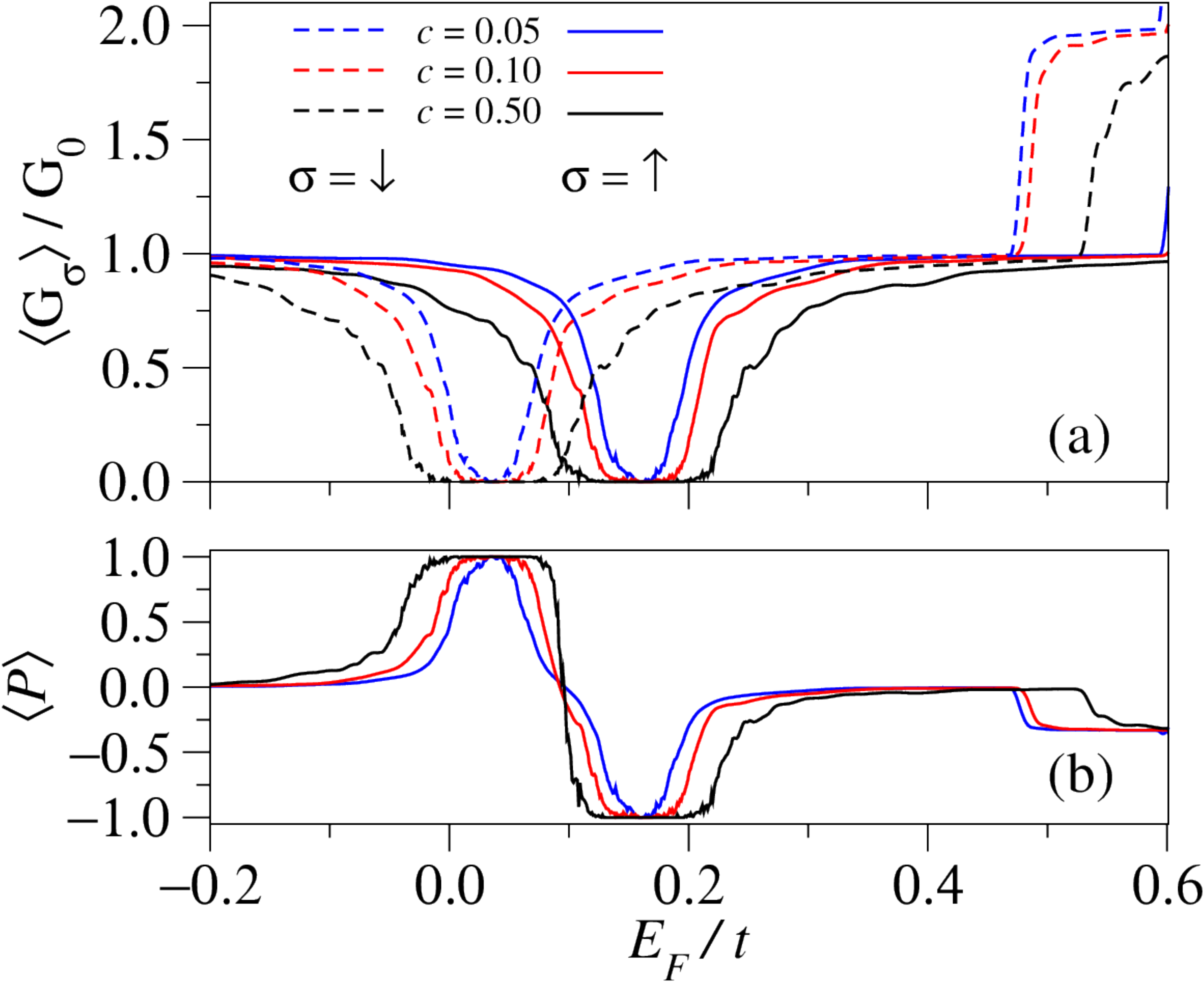}
\caption{(Color online) (a)~Spin-up (solid lines) and spin-down (dashed lines) conductances as a function of the Fermi energy of a silicene nanoribbon when $M=\unit{100}\milli\electronvolt$, averaged over $100$ realizations of random samples with adatom concentration $c=0.05$, $0.10$ and $0.50$. The energy level of the adatom is $\varepsilon_\mathrm{ad}=0.10t$ and the hopping parameter $t_\mathrm{ad}=0.25t$ with $t=\unit{1.6}\electronvolt$. 
(b)~Average conductance polarization as defined by Eq.~(\ref{eq:06}).}
\label{fig5}
\end{figure}
As mentioned before, half-metallicity of the system arises as a consequence of the spin-splitting induced by the ferromagnetic insulator. We describe  the  splitting  by  the  parameter $M$ in Eq.~(\ref{eq:04}). Ab initio calculations obtain values of the order of $\unit{100-200}\milli\electronvolt$ for graphene in close proximity to chalcogenides (EuO and EuS).\cite{Hallal17} Unfortunately, no similar calculations have been carried out in silicene yet and the magnitude of the parameter $M$ is largely unknown. In our numerical simulations we take a moderate value $M=\unit{100}\milli\electronvolt$ to be on the safe side although higher values are expected to result in better performance. 

In Fig.~\ref{fig5}(a) we show the resulting spin-dependent conductance $\langle G_{\sigma}\rangle$ ($\sigma=\uparrow,\downarrow$) as a function of the Fermi energy $E_F$ for  both spin directions, averaged over $100$ realizations of random samples with $c=0.05$, $c=0.10$ and $c=0.50$. The shape of the conductance curve is essentially the same for both spin orientations but blue and red shifted for spin up and spin down by an amount $M$, respectively. Fano anti-resonances are wider than those shown in Fig.~\ref{fig2} for the same adatom concentration because $t_\mathrm{ad}$ is now larger. We can then take advantage of the abrupt profile of the conductance curves to generate spin-polarized electric current through the silicene nanoribbon, as deduced from the polarization shown in Fig.~\ref{fig5}(b). When the Fermi level lies in the vicinity of $\varepsilon_\mathrm{ad}-M$ the electric current will be fully spin-down polarized. Similarly, when Fermi level approaches $\varepsilon_\mathrm{ad}+M$ the electric current will become fully spin-up polarized. By increasing the adatom concentration, the plateaus of the polarization around $\varepsilon_\mathrm{ad}\pm M$ can be made wide enough to ensure thermal stability in device applications, that is, keeping the width of the anti-resonance larger than $k_BT$.

\section{Conclusions}

We have studied the electrical conductance of narrow silicene nanoribbons\cite{Padova10} in close proximity to a ferromagnetic insulator. The ferromagnet induces a spin-splitting of the energy levels of the silicene nanoribbon. The magnitude of the splitting in silicene is still unknown and the we have used the same value found in graphene grown on chalcogenides\cite{Hallal17} to illustrate the phenomenon. We have also investigated the impact of a random distribution of adatoms adsorbed on the nanoribbon. The linear conductance shows clear signatures of the electronic Fano effect due to the coupling of the localized states at the adatoms and the continuum of propagating states in the nanoribbon. The Fano anti-resonance becomes spin-dependent due to the proximity exchange interaction between the itinerant electrons and the magnetic ions of the ferromagnet. The effect is robust and can be tuned by setting parameters like the adatom concentration. Moreover, it could be used to generate spin-polarized currents for real-world applications in spintronics.

\acknowledgments

C.~N. and F.~D-A. thank the Theoretical Physics Group of the University of Warwick for the warm hospitality. Work at Madrid has been supported by MINECO under Grants MAT2013-46308 and MAT2016-75955. L.~R and P.~A.~O. acknowledge FONDECYT Grants 1140388 and 1140571, respectively. C.~N. is grateful for the funding of scholarship CONICYT-Chile and DGIP-UTFSM. UK research data statement: All data accompanying this publication are directly available within the publication.

\end{document}